\documentclass[aps,prd,twocolumn,superscriptaddress,nofootinbib,10pt,reprint]{revtex4-2}

\usepackage{newtxtext,newtxmath}
\usepackage{amsmath,bm,mathtools}
\usepackage{graphicx}
\usepackage{booktabs,multirow,makecell}
\usepackage[dvipsnames]{xcolor}
\usepackage{hyperref}
\usepackage{url}
\usepackage{placeins}
\usepackage{microtype}
\usepackage{orcidlink}

\hypersetup{
	colorlinks=true,
	linkcolor=blue,
	citecolor=ForestGreen,
	urlcolor=blue
}

\newcommand{\rhom}{\ensuremath{\rho_{\rm m}}}
\newcommand{\rhode}{\ensuremath{\rho_{\rm de}}}
\newcommand{\rhor}{\ensuremath{\rho_{\rm r}}}

\newcommand{\paperfigure}[4][\linewidth]{%
	\begin{figure*}[t]
		\centering
		\IfFileExists{#2}{%
			\includegraphics[width=#1]{#2}%
		}{%
			\fbox{\parbox[c][0.18\textheight][c]{0.90\textwidth}{\centering
					\textbf{Figure placeholder}\\[1mm]\texttt{\detokenize{#2}}\\[1mm]
					Copy the final vector PDF into the \texttt{figures/} directory.}}%
		}
		\caption{#3}
		\label{#4}
	\end{figure*}
}

\begin{document}
	
	\title{Does DESI point towards transient cosmic acceleration?}
	
	\author{Muhammad Ahmad\orcidlink{0009-0000-9635-5065}}
	\email{ahmad@xao.ac.cn}
	\affiliation{State Key Laboratory of Radio Astronomy and Technology, Xinjiang Astronomical Observatory, CAS, 150 Science 1-Street, Urumqi, Xinjiang, 830011, P. R. China}
	\affiliation{School of Astronomy and Space Science, University of Chinese Academy of Sciences, No.1 Yanqihu East Road, Beijing 101408, China}
	
	\author{Saddam Hussain\orcidlink{0000-0001-6173-6140}}
	\email{saddamh@zjut.edu.cn; (Corresponding Author)}
	\affiliation{Institute for Theoretical Physics and Cosmology, Zhejiang University of Technology, Hangzhou 310023, China}

    \author{Ayodeji Ibitoye\orcidlink{0000-0002-0966-8598}}
	\email{ayodeji.ibitoye@gtiit.ac.cn}
	\affiliation{Department of Physics, Guangdong Technion – Israel Institute of Technology, Shantou, Guangdong 515063, People’s Republic of China}
	\affiliation{Centre for Space Research, North-West University, Potchefstroom 2520, South Africa }
	\affiliation{Department of Physics and Electronics, Adekunle Ajasin University, P. M. B. 001, Akungba-Akoko, Ondo State, Nigeria}
   
    \author{Yi'an Zhao}
	\email{zhaoyian@xao.ac.cn}
	\affiliation{State Key Laboratory of Radio Astronomy and Technology, Xinjiang Astronomical Observatory, CAS, 150 Science 1-Street, Urumqi, Xinjiang, 830011, P. R. China}
    \affiliation{School of Astronomy and Space Science, University of Chinese Academy of Sciences, No.1 Yanqihu East Road, Beijing 101408, China}      
   
   \author{Ming Zhang}
	\email{zhangm@xao.ac.cn}
	\affiliation{State Key Laboratory of Radio Astronomy and Technology, Xinjiang Astronomical Observatory, CAS, 150 Science 1-Street, Urumqi, Xinjiang, 830011, P. R. China}
	\affiliation{School of Astronomy and Space Science, University of Chinese Academy of Sciences, No.1 Yanqihu East Road, Beijing 101408, China}
	\affiliation{Xinjiang Key Laboratory of Radio Astrophysics, 150 Science 1-Street, Urumqi 830011, China}
    
	\author{Lang Cui\orcidlink{0000-0003-0721-5509}}
	\email{cuilang@xao.ac.cn; (Corresponding Author)}
	\affiliation{State Key Laboratory of Radio Astronomy and Technology, Xinjiang Astronomical Observatory, CAS, 150 Science 1-Street, Urumqi, Xinjiang, 830011, P. R. China}
	\affiliation{School of Astronomy and Space Science, University of Chinese Academy of Sciences, No.1 Yanqihu East Road, Beijing 101408, China}
	\affiliation{Xinjiang Key Laboratory of Radio Astrophysics, 150 Science 1-Street, Urumqi 830011, China}
	\date{\today}
	
	\begin{abstract}
		We investigate an oscillatory equation of state for dark energy that oscillates around the mean value $-1$ and includes a source term proportional to the cube of the number of e-folds. Due to this source term, the oscillation amplitude shifts towards the $w_{\rm de}<-1$ regime at higher redshifts, whereas in the future, the source term drives the oscillation amplitude towards the $w_{\rm de}>-1$ regime. As the equation of state enters the $w_{\rm de}>-1$ regime in the future, the dark energy density dilutes, causing the Hubble parameter to decrease asymptotically towards zero, so that the present acceleration is transient: the Universe transitions to decelerated expansion and approaches a static state as $a\to\infty$, without ever contracting. We compare this equation of state with BellDE, CPL and flat $\Lambda$CDM. We confront the model with a combination of datasets, including Planck, DESI BAO and three different supernova compilations: Pantheon+, Union3 and DES-Dovekie. We compare the log-evidence of each model with that of $\Lambda$CDM. The dynamical dark energy models yield $H_0$ values in the range $66.0^{+0.80}_{-0.79}$--$67.84^{+0.69}_{-0.74}\,{\rm km\,s^{-1}\,Mpc^{-1}}$, with matter density values in the range $\Omega_{\rm m}=0.3067^{+0.0073}_{-0.0072}$--$0.3271^{+0.0083}_{-0.0080}$ across all the combinations. These values are consistent with those of the baseline model, apart from a mild upward shift in $\Omega_{\rm m}$ for the Union3 combination. Among all the models, BellDE yields comparatively stronger Bayesian evidence than the other models, with weak evidence for the Pantheon+ compilation and moderate evidence for the Union3 and DES-Dovekie compilations. The sourced oscillatory model yields inconclusive evidence for Pantheon + and Union3, while weak evidence for DES-Dovekie relative to $\Lambda$CDM.
	\end{abstract}
	
	\maketitle

	\section{Introduction}
	\label{sec:introduction}
	
	The discovery of the accelerated expansion of the Universe from Type Ia supernova observations established one of the most challenging problems in modern cosmology \cite{SupernovaSearchTeam:1998fmf,SupernovaCosmologyProject:1998vns}. Within the standard model of cosmology, this acceleration is described by a positive cosmological constant, $\Lambda$, with an equation of state $w_{\Lambda}=-1$, together with a pressureless, electromagnetically neutral dark component known as cold dark matter (CDM). The resulting spatially flat $\Lambda$CDM model provides a successful description of the cosmic microwave background (CMB) anisotropies and is broadly consistent with baryon acoustic oscillation (BAO) measurements, large-scale structure observations, and Type Ia supernova measurements \cite{Planck:2018vyg,Eisenstein:2005sbt,SDSS:2005xqv,Primack:1997av,DelPopolo:2007dna,Diao:2023tor,Mina:2020eik,Blumenthal:1984bp,Baccigalupi:2006ww}. However, its phenomenological success has not yet resolved the underlying nature of the accelerated expansion. In particular, the enormous discrepancy between the observed vacuum energy density and simple quantum field theory expectations, together with the coincidence that the dark energy and matter densities become comparable only at relatively recent cosmic times, continues to motivate alternatives to a strictly constant vacuum energy \cite{Weinberg:1988cp,Sahni:1999gb,Sahni:2004ai,Frieman:2008sn,Copeland:2006wr}.
	
	A natural possibility is that the component responsible for late-time acceleration evolves with time. Such behavior arises in a broad class of theoretical constructions, including canonical scalar-field models \cite{Ratra:1987rm,Zlatev:1998tr,Tsujikawa:2013fta}, non-canonical scalar-field models \cite{Chiba:1999ka,Armendariz-Picon:2000nqq,Armendariz-Picon:2000ulo,Hussain:2024qrd}, interactions between the dark sectors \cite{Amendola:1999er,Farrar:2003uw,Hussain:2025uye}, and modifications of the gravitational sector \cite{Sotiriou:2008rp,Nojiri:2006ri,Cai:2015emx}. Although these scenarios have different underlying physical origins, much of their phenomenology can be characterized, at the background level, by an effective dark energy equation of state,
	\begin{equation*}
		w_{\rm de}(z)=\frac{p_{\rm de}(z)}{\rho_{\rm de}(z)} .
	\end{equation*}
	Beyond its theoretical motivation, the dynamical nature of dark energy has received renewed attention following recent observations from the Dark Energy Spectroscopic Instrument (DESI) \cite{DESI:2024mwx,DESI:2024kob}. When the observational measurements are analyzed using the two-parameter Chevallier--Polarski--Linder (CPL) equation-of-state parametrization, they yield a $2.6$--$3.9\sigma$ preference for dynamical dark energy over a cosmological constant \cite{DESI:2025zgx,DESI:2025wyn,Linder:2002et} depending on supernova compilations. The CPL parametrization, constructed from a Taylor expansion around $a=1$, provides a simple phenomenological framework for probing the dynamical nature of dark energy. However, its functional form can lead to unphysical behavior when extrapolated into the far future, limiting its utility for predicting the future evolution of the Universe. Moreover, it cannot adequately capture non-monotonic evolution of the equation of state, which may be favoured by observational data \cite{DESI:2025wyn}. From an observational perspective, accounting for the variations in observational sample can help clarify the nature of dark energy and potentially identify physically viable theories of gravity on cosmological scales. Motivated by these limitations of the CPL parametrization, various alternative functional forms of the equation of state have been proposed that avoid divergences in the future and incorporate nonlinear behavior, allowing for non-monotonic evolution with cosmic time or redshift \cite{Li:2011dr,Yang:2017alx,Mehrabi:2018oke,Escamilla:2021uoj,Ormondroyd:2025exu,Ormondroyd:2025iaf,Hussain:2025nqy,Rezaei:2024vtg,Kessler:2025kju,Escamilla:2024fzq,Pan:2019brc,Li:2019yem,Castillo-Santos:2022yoi,Yang:2021flj,Escamilla:2023oce,Ormondroyd:2025exu,Hussain2026DHO,Hussain:2026srf}.
	
	In this paper, we propose an equation of state that exhibits simple harmonic oscillations with frequency $f$ around the mean value $w_{\rm m}$ and satisfies the differential equation
	\begin{equation}
		\frac{d^2 w_{\rm de}}{dN^2} + f^2 (w_{\rm de} - w_{\rm m}) = \Omega_{\rm m0} N^3\ , \label{sourced_oscillating}
	\end{equation}
	where $N=\ln(a)$ is the logarithm of the scale factor $a$ and serves as the time variable. The evolution of the equation of state is determined by the initial conditions $w(N=0)=w_0$ and $dw/dN|_{N=0}=w_1$. The oscillatory system is driven by a source term proportional to the cube of the number of e-folds, with the proportionality constant set to the present-day matter density parameter, $\Omega_{m0}$. The solution to this differential equation is
	\begin{multline}
		w_{\rm de}(N) = (w_0 - w_{\rm m}) \cos(f N) + \frac{(6 \Omega_{\rm m0} + f^4 w_1)\sin(f N)}{f^5} \\  + w_{\rm m} + \dfrac{\Omega_{\rm m0}N^3}{f^2} - \dfrac{6\Omega_{\rm m0} N}{f^4}\ . \label{sol_so}
	\end{multline}
	In the absence of the source term, the equation of state exhibits simple harmonic oscillations around the mean value $w_{\rm m}=-1$ with frequency $f$. The source term introduces two odd powers of the time variable, $N^3$ and $N$, and shifts the coefficient of the sine term, thereby modifying the locations of the oscillation extrema. As the system evolves into the past ($N<0$), the cubic term becomes increasingly important as $N$ increases to higher negative values, driving the equation of state towards more negative values. Consequently, the model can exhibit phantom behavior. As the system evolves into the future ($N>0$), the equation of state tends to increase towards positive values. However, the negative linear term in $N$ moderates the growth of the cubic contribution over the range $-2<N<2$. Thus, at low frequencies, the source term can modify the oscillation amplitude and produce dynamics that differ from $\Lambda$CDM and CPL, providing a useful framework for testing alternatives to dynamical dark energy and modified theories of gravity. We refer to this parametrization as the sourced oscillatory (SO) model of dark energy.
	
	In addition to the SO model, we compare our model with the BellDE equation of state introduced in Ref.~\cite{Hussain:2025nqy}, which exhibits multiple crossings of $w_{\rm de}=-1$ and can be systematically extrapolated into the future. We compare both models with CPL and $\Lambda$CDM using various combinations of observational datasets. We adopt Planck observations \cite{Planck:2018vyg,Planck:2018nkj,Aghanim:2019ame,Aghanim:2018oex} and DESI BAO measurements \cite{DESI:2025zgx} as our baseline datasets, and incorporate three distinct supernova compilations separately: Pantheon+, Union3, and DES-Dovekie \cite{Brout:2022vxf,Rubin:2023jdq,DES:2025sig}. The SO model yields $H_0\simeq 66.1 - 67.8$ km/s/Mpc depending on the supernova compilation similar to BellDE and CPL. In contrast, $\Lambda$CDM yields $H_0\sim 68.30\pm0.29$ km/s/Mpc, exhibiting a tension of nearly $4\sigma$ with the value inferred by SH0ES \cite{Brout:2022vxf}. We emphasize that none of the dynamical models considered here alleviates this tension: all of them yield lower $H_0$ values than $\Lambda$CDM. Beyond the $H_0$ discrepancy, the present-day equation-of-state parameter remains in the quintessence regime for Pantheon+ and Union3, with $w_0=-0.945^{+0.096}_{-0.111}$ and $w_0=-0.778^{+0.069}_{-0.069}$, respectively, whereas its central value lies in the phantom regime for DES-Dovekie, with $w_0=-1.089^{+0.242}_{-0.376}$. Furthermore, BellDE yields a present-day equation-of-state value close to that of CPL and a present-day matter density parameter of $\Omega_{m0}\sim0.31$, whereas the SO model yields $\Omega_{m0}\sim0.30$, close to the $\Lambda$CDM value except for the Union3 compilation. Thus, the SO model is dynamically driven, and the constraints obtained for it remain close to those of both CPL and $\Lambda$CDM, while its oscillatory phase differs across the supernova compilations. This compilation dependence itself provides a useful diagnostic of the systematics affecting low- and moderate-redshift supernova samples, which monotonically varying equation-of-state models cannot capture. 
	
	The Bayesian log-evidence for BellDE indicates moderate support, with $\Delta\log Z\sim3.6$ relative to $\Lambda$CDM for both Union3 and DES-Dovekie, whereas the SO model yields inconclusive evidence relative to the baseline model. Across the dynamical models and supernova compilations, the matter clustering amplitude lie in the range $S_8\simeq 0.814$ -- $0.833$, while $\Lambda$CDM yields a slightly lower value of $S_8\sim0.811$, indicating a small difference between the dynamical models and the cosmological constant scenario.
	
	The remainder of the paper is organized as follows. Section~\ref{sec:background} introduces the background dynamics and the four cosmological models. Section~\ref{sec:data_analysis} describes the observational data and numerical analysis. Section~\ref{sec:results} presents the cosmological constraints and discusses their observational signatures. Finally, we conclude in Section~\ref{sec:conclusions}.

	\section{Background dynamics and the models}
	\label{sec:background}
	
	Assuming that the Universe is homogeneous, isotropic, and spatially flat on the largest scales, the corresponding line element is given by
	\begin{equation}
		ds^2=-dt^2+a^2(t)d\boldsymbol{x}^2 ,
	\end{equation}
	where $a(t)$ is the scale factor. The first Friedmann equation corresponding to this metric reads
	\begin{equation}
		3H^2 = \kappa^2 \left(\rhode + \rhom + \rhor \right)
		\label{eq:friedmann}
	\end{equation}
	where the evolution of the Hubble parameter $H$ is governed by the energy content of the Universe, including the radiation energy density $\rhor$, the pressureless matter density $\rhom$, and the dark energy density $\rhode$. Here, $\kappa$ is the inverse of the reduced Planck mass, $M_{\rm pl}$. All components are assumed to be minimally coupled to the gravitational sector and thus independently satisfy the continuity equation
	\begin{equation}
		\dot{\rho}_i + 3H(1+w_{i})\rho_i=0 ,
		\label{eq:continuity}
	\end{equation}
	where the equation-of-state parameter is defined as $w_i\equiv\frac{p_i}{\rho_i}$. The energy densities of matter and radiation evolve with the redshift $z$ as
	\begin{equation}
		\rhor = \rho_{\rm r0} (1+z)^{4}, \ \ \rho_{\rm m} = \rho_{\rm m0} (1+z)^{3},
	\end{equation}
	whereas, using the continuity equation, the dark energy density can be expressed as
	\begin{equation}
		\ln \bigg(\frac{\rho_{\rm de}(N)}
		{\rho_{{\rm de},0}} \bigg)
		=-3\int_0^N
		\left(1+w_{\rm de}(\widetilde N)\right)
		d\widetilde N\ .
		\label{eq:rho_de_general}
	\end{equation}
	To describe the dynamics, we consider three distinct parametrizations of the dark energy equation of state:
	\begin{eqnarray}
		\text{\bf Bell:}  \ w_{\rm de}(z) &= & w_1^{\rm Bell} +
		\left(w_0^{\rm Bell} 	- w_1^{\rm Bell}
		\right) \exp\!\left[ -\left(\frac{z-z_t}{\Delta}
		\right)^2\right].
		\label{eq:BellDE}\\
		\text{\bf SO:} \ 	\frac{d^2 w_{\rm de}}{dN^2} & = & - f^2 (w_{\rm de} - w_{\rm m}) + \Omega_{\rm m0}\  N^3\ ,\\
		\text{\bf CPL:} \ w_{\rm de}(a) &= & w_0+w_a (1-a) \ .
		\label{eq:CPL}
	\end{eqnarray}
	Here, $z$ denotes the redshift, and $N=-\ln(1+z)$ is the number of e-folds.
	
	\subsection{Bell-shaped dark energy}
	\label{subsec:BellDE_theory}
	
	The Gaussian functional form of the dark energy equation of state in Eq.~\eqref{eq:BellDE} depends on four free parameters, $\{w_0^{\rm Bell}, w_1^{\rm Bell}, z_t, \Delta\}$. This phenomenological parametrization was introduced in Ref.~\cite{Hussain:2025nqy} to investigate whether dark energy remains phantom at higher redshifts or can transition into the quintessence regime. Consequently, a localized, redshift-dependent form of the equation of state was adopted.
	
\begin{figure*}[t]
		\centering
		\includegraphics[scale=0.6]{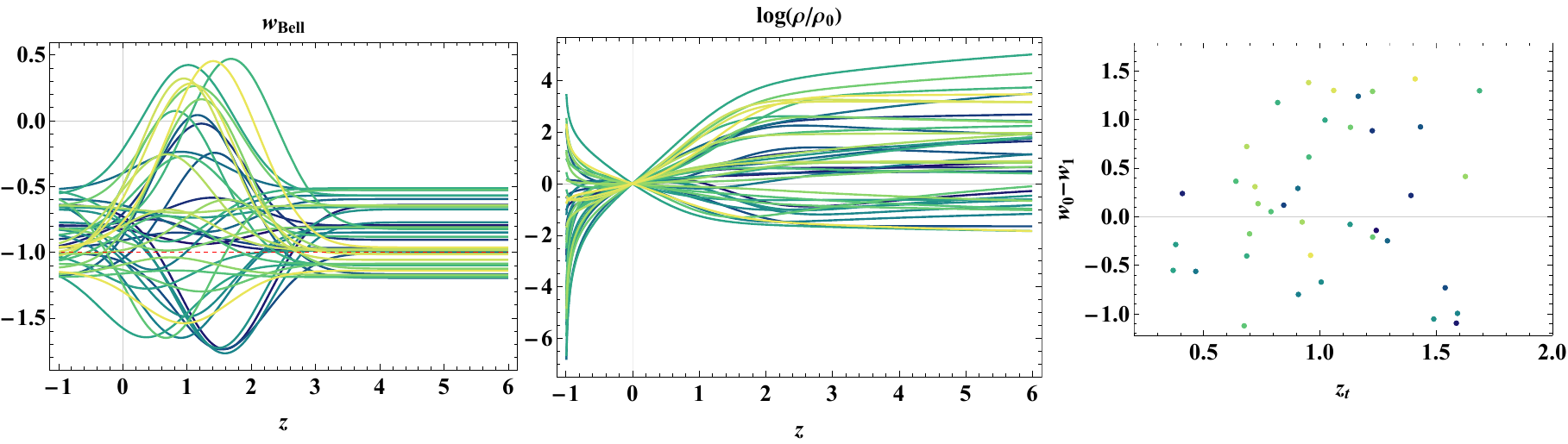}
		\caption{
			Variation of the BellDE equation of state and the corresponding dark energy density as functions of redshift $z$, for a fixed $\Delta=1.0$, which controls the width of the Gaussian function. The curves illustrate different values of $w_0\in[-2,0.5]$, $w_1\in[-1.2,-0.8]$, and $z_t\in[0.3,1.7]$. Depending on the value of $(w_0-w_1)$, the Gaussian excursion extends towards more negative values, corresponding to stronger phantom behavior, or towards less negative values, corresponding to the quintessence regime. For each variation of the equation-of-state curve, the dark energy density either increases or decreases with increasing redshift.
		}
		\label{fig:BellDE_theory}
	\end{figure*}

	The evolution of the equation of state is shown in Fig.~\ref{fig:BellDE_theory}. At redshifts far from $z_t$, the equation of state asymptotically approaches $w_1^{\rm Bell}$ in both the distant past and the distant future, whereas at $z=z_t$, it reaches $w_{\rm de}(z_t)=w_0^{\rm Bell}$. The parameter $z_t$ determines the redshift at which the excursion is centered, while $\Delta$ controls its characteristic width. The corresponding dark energy density evolves differently depending on the behavior of the equation of state. For trajectories in which the equation of state falls below $-1$ around $z\gtrsim1$ and remains below this value, the dark energy density decreases with increase in redshift. Conversely, for trajectories in which the equation of state remains above $-1$, the dark energy density increases with increase in redshift. However, its magnitude may not remain subdominant to the matter density, and such solutions may therefore influence the matter-dominated epoch of the Universe. Accordingly, in our MCMC analysis, we vary all the parameters over sufficiently broad ranges to allow for the full range of possible solutions permitted by the model. The first derivative of the equation of state is given by
	\begin{equation}
		\frac{dw_{\rm de}}{dz} 	=
		-\frac{2(z-z_t)}{\Delta^2}
		\left(
		w_0^{\rm Bell}
		-
		w_1^{\rm Bell}
		\right)
		\exp\!\left[
		-\left(
		\frac{z-z_t}{\Delta}
		\right)^2
		\right],
		\label{eq:BellDE_derivative}
	\end{equation}
	which vanishes at $z=z_t$ and asymptotically approaches zero for $|z-z_t|\gg\Delta$, indicating negligible variation far from $z_t$. The present-day value of the equation of state cannot be directly identified with $w_0^{\rm Bell}$; instead, it must be evaluated as
	\begin{equation}
		w_{\rm de}(0)	=w_1^{\rm Bell}
		+	\left(	w_0^{\rm Bell} 	- 	w_1^{\rm Bell} 	\right)
		\exp\!\left[	-\left( 	\frac{z_t}{\Delta} 	\right)^2 \right].
		\label{eq:BellDE_present}
	\end{equation}
	
	\subsection{Sourced Oscillation model}
	\label{subsec:SO_theory}
	
\begin{figure*}[t]
		\centering
		\includegraphics[scale=0.6]
		{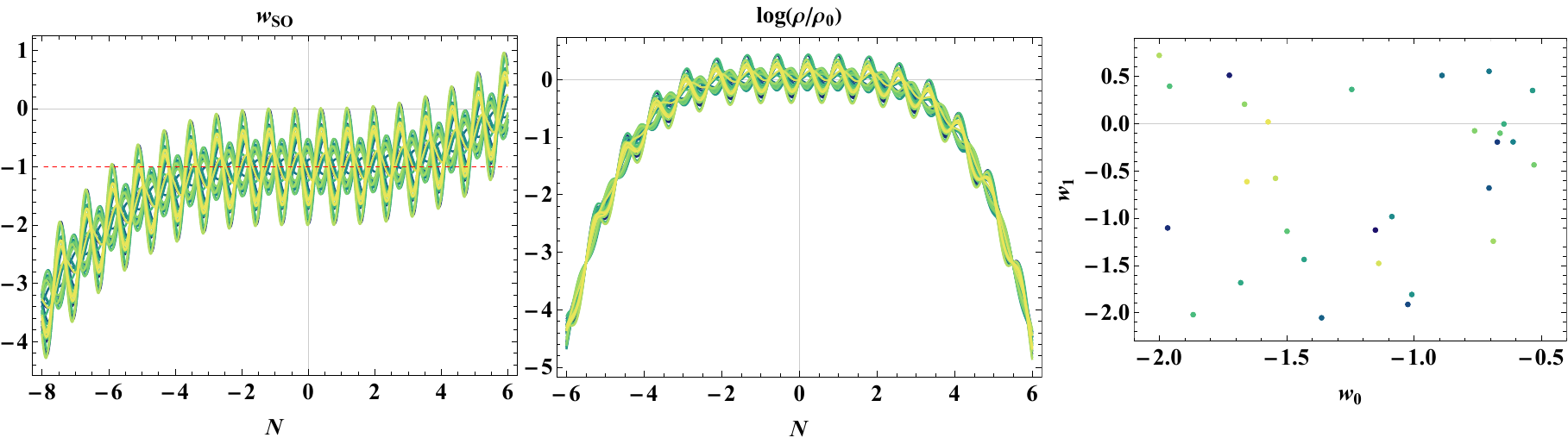}
		\caption{
			Variation of $w_{\rm de}(N)$ with respect to the number of e-folds, $N=-\ln(1+z)$, for fixed $f=8.0$ and $w_{\rm m}=-1$. We vary the initial conditions $w_0\in[-2.0,-0.5]$ and $w_1\in[-2.0,-0.3]$, with $\Omega_{\rm m0}=0.3$. Negative values of $N$ correspond to the past evolution, whereas $N>0$ describes the future evolution of the system.
		}
		\label{fig:SO_theory}
	\end{figure*}

	The governing equation for this equation of state is given in Eq.~\eqref{sourced_oscillating}, and its corresponding solution is presented in Eq.~\eqref{sol_so}. The model has three free parameters: two initial conditions required to solve the differential equation, $\{w_0, w_1\}$, and the oscillation frequency $f$. In general, there is no fundamental reason to choose the proportionality constant of the source term to be the present-day matter density parameter. Hence, one is free to choose any constant, provided it is positive. A negative value can drive the peaks of the oscillation towards $w_{\rm de}>-1$ and cause the dark energy density to grow towards the past, potentially altering the baryon and radiation densities and rendering the model physically non-viable. The evolution of the equation of state and the corresponding energy density is illustrated in Fig.~\ref{fig:SO_theory}. We plot the variation by fixing $f=8.0$ and $\Omega_{\rm m0}=0.3$, with the oscillations centered around the mean value $w_{\rm m}=-1$. Due to the source term proportional to $N^3$, the equation of state oscillates at low redshifts with minimal influence from the source term. At large negative values of $N$, corresponding to the distant past, the mean of the oscillations shifts towards more negative values ($<-1$), resulting in a rapid decrease in the corresponding dark energy density. As the system evolves into the future, positive values of $N$ drive the equation of state (EOS) gradually towards the non-accelerating regime, causing the corresponding dark energy density to dilute rapidly. This eventually drives the Hubble parameter asymptotically toward zero. Thus, in a spatially flat FLRW universe with a non-negative total energy density, $H\to0$ implies $\dot a\to0$. The model therefore predicts a decelerating future expansion following a transient epoch of accelerated expansion. Although the model currently lacks a fundamental microscopic motivation, it naturally describes transient cosmic acceleration, in contrast to the eternal de Sitter future predicted by the cosmological constant. Thus, the model exhibits behavior similar to that of CPL while allowing a consistent extrapolation into the future.
	
	It is also worth noting that the source-term normalization $\Omega_{\rm m0}$, can also be set to some arbitrary parameter lets call it $\lambda$. However, if we look at the solution of EOS in Eq.~\eqref{sol_so}, the normalization always appears in combination of other parameters such as $\lambda/f^2$, $\lambda/f^4$ governing the drift of the oscillation mean, and $\lambda/f^5$ correcting the sine amplitude. The data therefore constrain these combinations rather than $\lambda$ and $f$ independently. Thus, fixing $\lambda=\Omega_{\rm m0}$, break the degeneracy between the parameters and eventually the data constraint $f$ that in another words should make the normalization physically unchanged. Therefore, the choice of normalization constant should be understood as a convention rather than the physical assumption.

	For the MCMC analysis and comparison of the statistical evidence with CPL, we restrict the number of free parameters to $\{w_0, f\}$, fixing $w_1=0.0$ and taking the source-term proportionality constant to be $\Omega_{\rm m0}$. This choice is motivated by the structure of the solution given in Eq.~\eqref{sol_so}, where the coefficients of the trigonometric functions can be expressed as
		\begin{equation}
			A \sin\delta = w_0 - w_{\rm m},\quad A \cos\delta = \frac{(6 \Omega_{\rm m0} + f^4 w_1)}{f^5},
		\end{equation}
		from which the phase of the oscillation is given by
		\begin{equation}
			\delta = \tan^{-1} \left(\dfrac{(w_0 - w_{\rm m})f^5}{(6 \Omega_{\rm m0} + f^4 w_1)}\right),
		\end{equation}
		while the amplitude of the oscillation is given by
		\begin{equation}
			A^2 = \left(w_0 - w_{\rm m}\right)^2 + \left(\frac{(6 \Omega_{\rm m0} + f^4 w_1)}{f^5}\right)^2.
		\end{equation}
		With this the equation of state in Eq.~\eqref{sol_so} becomes
		\begin{equation}
			w_{\rm de}(N) = w_{\rm m} + A \sin(f N + \delta) + \dfrac{\Omega_{\rm m0}N^3}{f^2} - \dfrac{6\Omega_{\rm m0} N}{f^4}\ .
		\end{equation}
		For $f\gg1$, the phase and amplitude become
		\begin{equation}
			\delta \sim \tan^{-1}\left(\frac{(w_0 - w_{\rm m})f}{w_1}\right),\quad
			A \sim \left(w_0 - w_{\rm m}\right).
		\end{equation}
		These expressions show that $w_1$ enters through a combination of $w_0$ and $f$, introducing a degeneracy among these parameters from the perspective of data analysis. Fixing $w_1$ therefore removes a degeneracy between parameters that current observations cannot independently constrain. Thus, similar to fixing the normalization parameter, fixing $w_1$ is a phase convention that does not remove any direction that the data could independently constraint. Therefore, the overall dynamics can be effectively evaluated by sampling $w_0$ and $f$.

	\section{Methodology and Data Analysis}
	\label{sec:data_analysis}

	We implement the models, along with $\Lambda$CDM, by modifying the \texttt{CLASS} code and perform parameter inference using the publicly available \texttt{COBAYA} Python pipeline \cite{Class_code,Torrado:2020dgo}. The converged chains are subsequently analyzed using \texttt{GetDist} \cite{Lewis:2019xzd} to obtain the marginalized parameter distributions at the 68\% and 95\% confidence levels. Throughout the analysis, we assume that the dark energy fluid does not cluster and, hence, set its corresponding sound speed to $c_{\rm s}^2=1$. Because $w_{\rm de}(z)$ crosses $-1$ multiple times in parts of the parameter space, dark-energy perturbations are treated within the parameterized post-Friedmann (PPF) framework as implemented in \texttt{CLASS}~\cite{Fang:2008sn}, with vanishing anisotropic stress. This prescription yields smooth, per-cent-level deviations in the matter and CMB power spectra, as shown explicitly in Fig.~\ref{fig:cmb_matter_power}.

	$\bullet$ We use the Planck 2018 high-$\ell$ TT, TE, and EE likelihoods, together with the low-$\ell$ temperature and $E$-mode polarization likelihoods \cite{Aghanim:2019ame,Planck:2018vyg}. We also include the Planck 2018 CMB lensing likelihood based on the reconstructed lensing potential \cite{Aghanim:2018oex}.
	
	$\bullet$ For the baryon acoustic oscillation (BAO) measurements, we use the DESI Data Release 2 (DR2) \cite{DESI:2025zgx}. The effective redshifts entering the likelihood span approximately $0.295\leq z_{\rm eff}\leq2.330$, including the high-redshift Ly$\alpha$ forest measurements. Depending on the tracer, DESI provides measurements of the volume-averaged distance $D_V/r_d$, the comoving angular diameter distance $D_M/r_d$, and the comoving Hubble distance $D_H/r_d$, all normalized by the sound horizon $r_d$ at the baryon drag epoch.
	
	$\bullet$ We include three distinct catalogs of Type Ia supernovae (SNe Ia). Pantheon+ (PP) contains 1701 light curves from 1550 distinct SNe Ia and incorporates improved cross-calibration and treatment of systematic uncertainties \cite{Brout:2022vxf}. Union3 contains 2087 cosmologically useful SNe Ia analysed within the UNITY1.5 Bayesian framework \cite{Rubin:2023jdq}. DES-Dovekie corresponds to an updated reanalysis of the Dark Energy Survey five-year (DES-SN5YR), incorporating improved photometric cross-calibration, updated low-redshift calibration, and a retrained SALT3 light-curve model \cite{DES:2025sig}. For simplicity, we refer to this likelihood as DES-SN hereafter.
	
	Because of the overlap between the supernova samples and their different calibration methods, we analyze these samples independently in combination with the baseline data set, consisting of Planck and DESI measurements. Thus, the total likelihood $\mathcal{L}_{\rm tot}$ is given by
	\begin{equation}
		\begin{aligned}
			&\mathcal{L}_{\rm CMB} + \mathcal{L}_{\rm DESI} + \mathcal{L}_{\rm PP}, \\
			&\mathcal{L}_{\rm CMB} + \mathcal{L}_{\rm DESI} + \mathcal{L}_{\rm Union3}, \\
			&\mathcal{L}_{\rm CMB} + \mathcal{L}_{\rm DESI} + \mathcal{L}_{\rm DES-SN} \ .
		\end{aligned}
		\label{eq:data_combinations}
	\end{equation}
	In addition to the model parameters, we vary the nuisance and cosmological density parameters associated with the baryon and cold dark matter densities,
	\begin{equation}
		\left\{
		\Omega_{\rm b}h^2,\,
		\Omega_{\rm c}h^2,\,
		100\theta_s,\,
		\tau_{\rm reio},\,
		n_s,\,
		\ln(10^{10}A_s)
		\right\} \ ,
		\label{eq:base_parameters}
	\end{equation}
	where $h\equiv H_0/100$. We adopt broad uniform priors for each parameter, as summarized in Tab. \ref{tab:parameter_priors}. The prior ranges for the nuisance parameters are chosen according to the standard Planck likelihood prescription.
	\begin{table}[t]
		\centering
		\renewcommand{\arraystretch}{1.15}
		\begin{tabular}{lc}
			\hline\hline
			Parameter & Prior range \\
			\hline
			$\Omega_{\rm b}h^2$       & $[0.005,\,0.100]$ \\
			$\Omega_{\rm c}h^2$       & $[0.001,\,0.990]$ \\
			$100\theta_s$             & $[1.03,\,1.05]$ \\
			$\tau_{\rm reio}$         & $[0.01,\,0.80]$ \\
			$n_s$                     & $[0.80,\,1.20]$ \\
			$\ln(10^{10}A_s)$         & $[1.61,\,3.91]$ \\
			\hline
			$w_0^{\rm Bell}$          & $[-4,\,1]$ \\
			$w_1^{\rm Bell}$          & $[-3,\,-0.05]$ \\
			$z_t$                     & $[0,\,5]$ \\
			$\Delta$                  & $[0.05,\,5]$ \\
			\hline
			$w_{0,\rm SO}$            & $[-3,\,2]$ \\
			$f_{\rm SO}$              & $[0,\,30]$ \\
			$w_1$ (SO)					& $0$ \\
			\hline
			$w_0$ (CPL)               & $[-3,\,1]$ \\
			$w_a$ (CPL)               & $[-3,\,2]$ \\
			\hline\hline
		\end{tabular}
		\caption{The uniform prior ranges corresponding to the models.}
		\label{tab:parameter_priors}
	\end{table}
	
	\section{Results}
	\label{sec:results}
	
\begin{figure*}[t]
		\begin{tabular}{cc}
			{\bf CPL} & {\bf \boldmath $\Lambda$CDM}\\
			\includegraphics[scale=0.45]{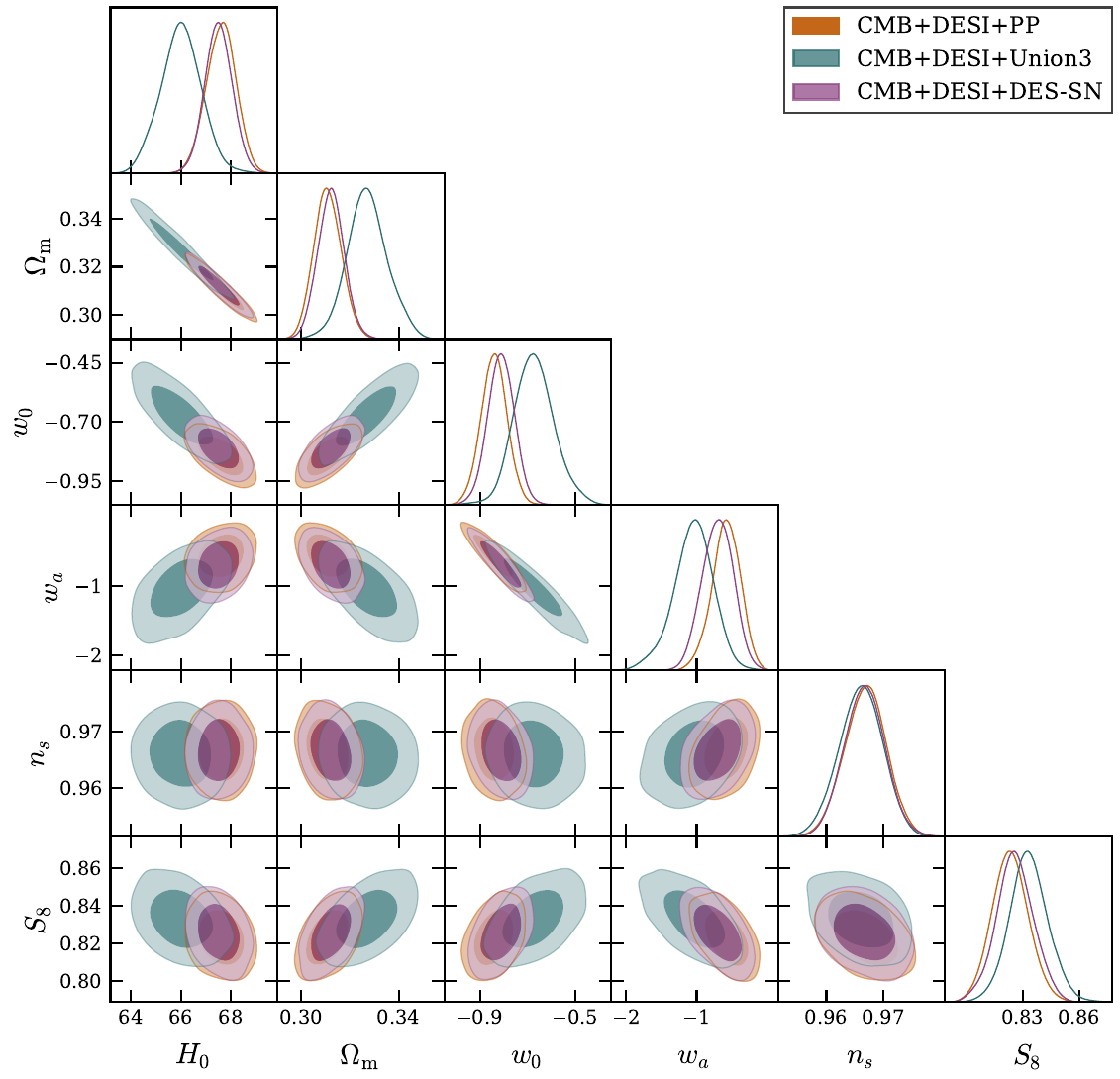} &
			\includegraphics[scale=0.45]{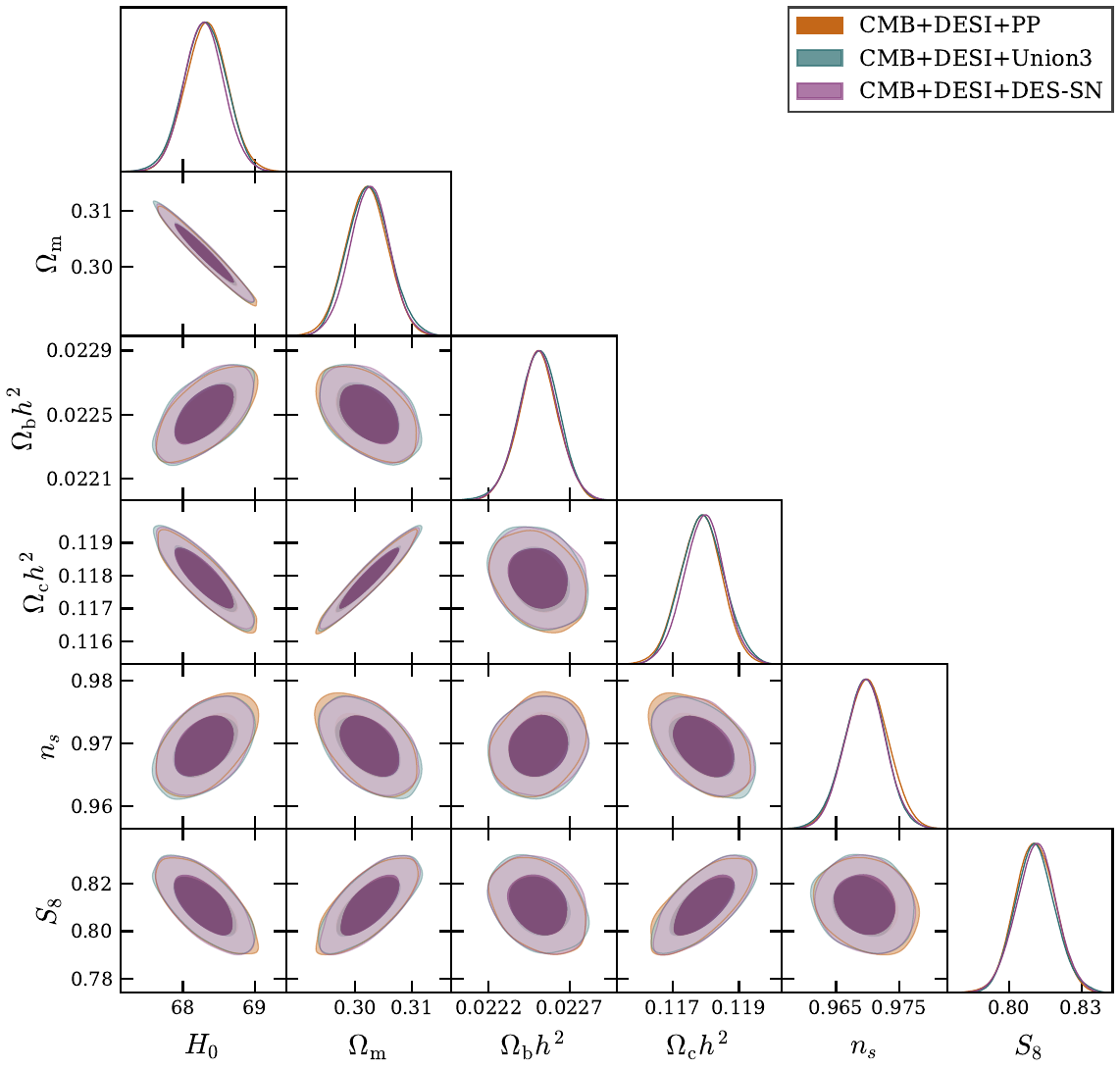}
		\end{tabular}
		\caption{Marginalized posterior distribution of the parameters corresponding to CPL and $\Lambda$CDM.}
		\label{fig:lcdm_cpl_posterior}
	\end{figure*}

	The marginalized parameter distribution for BellDE, SO, CPL, and flat $\Lambda$CDM are shown in Figs. \ref{fig:Bell_posterior}, and  \ref{fig:lcdm_cpl_posterior} and the corresponding constraints are summarized in Tab.~\ref{tab:combined_cosmo_constraints_68_fourmodel}. All quoted intervals correspond to the posterior means and central $68\%$ credible regions. We find that the parameters constrained by the CMB likelihood remain comparatively similar across all models. The remaining parameters, particularly the matter density $\Omega_{\rm m0}$ and $H_0$, show some variation across the different supernova compilations but remain consistent among all the dynamical dark energy models. However, a slight discrepancy in these parameters can be observed between $\Lambda$CDM and the dynamical dark energy models.
	
	\begin{table*}[t]
		\centering
		\footnotesize
		\renewcommand{\arraystretch}{1.10}
		\resizebox{\textwidth}{!}{%
			\setlength{\tabcolsep}{2.8pt}
			\begin{tabular}{lccccccccccc}
				\toprule
				\toprule
				Dataset & $\chi^2_{\min}$ & $H_0$ & $\Omega_{\rm m}$ & $w_{0}^{\rm Bell}$ & $w_{1}^{\rm Bell}$ & $z_t$ & $\Delta$ & $w_{0,\rm SO}$ & $f_{\rm SO}$ & $w_0$ & $w_a$ \\
				\midrule
				\multicolumn{12}{l}{\textbf{BellDE cosmology}} \\
				CMB+DESI+PP & $4183.939$ & $67.72_{-0.59}^{+0.60}$ & $0.3104_{-0.0058}^{+0.0057}$ & $-2.142_{-0.874}^{+0.712}$ & $-0.671_{-0.249}^{+0.307}$ & $3.584_{-1.029}^{+1.024}$ & $2.674_{-1.282}^{+1.358}$ & \textemdash & \textemdash & \textemdash & \textemdash \\
				CMB+DESI+Union3 & $2800.990$ & $66.73_{-0.78}^{+0.78}$ & $0.3201_{-0.0080}^{+0.0078}$ & $-2.261_{-0.711}^{+0.672}$ & $-0.527_{-0.282}^{+0.287}$ & $3.517_{-1.114}^{+1.070}$ & $2.693_{-1.140}^{+1.131}$ & \textemdash & \textemdash & \textemdash & \textemdash \\
				CMB+DESI+DES-SN & $4411.343$ & $67.64_{-0.54}^{+0.54}$ & $0.3111_{-0.0052}^{+0.0053}$ & $-2.110_{-0.800}^{+0.671}$ & $-0.605_{-0.282}^{+0.320}$ & $3.637_{-1.004}^{+0.987}$ & $2.867_{-1.151}^{+1.186}$ & \textemdash & \textemdash & \textemdash & \textemdash \\
				\midrule
				\multicolumn{12}{l}{\textbf{SO cosmology}} \\
				CMB+DESI+PP & $4187.107$ & $67.82_{-0.70}^{+0.63}$ & $0.3069_{-0.0065}^{+0.0070}$ & \textemdash & \textemdash & \textemdash & \textemdash & $-0.945_{-0.111}^{+0.096}$ & $11.291_{-7.951}^{+12.686}$ & \textemdash & \textemdash \\
				CMB+DESI+Union3 & $2802.140$ & $66.14_{-0.76}^{+0.75}$ & $0.3241_{-0.0082}^{+0.0082}$ & \textemdash & \textemdash & \textemdash & \textemdash & $-0.778_{-0.069}^{+0.069}$ & $3.992_{-0.609}^{+0.570}$ & \textemdash & \textemdash \\
				CMB+DESI+DES-SN & $4414.010$ & $67.84_{-0.74}^{+0.69}$ & $0.3067_{-0.0072}^{+0.0073}$ & \textemdash & \textemdash & \textemdash & \textemdash & $-1.089_{-0.376}^{+0.242}$ & $13.237_{-9.569}^{+14.923}$ & \textemdash & \textemdash \\
				\midrule
				\multicolumn{12}{l}{\textbf{CPL cosmology}} \\
				CMB+DESI+PP & $4186.880$ & $67.60_{-0.59}^{+0.58}$ & $0.3108_{-0.0056}^{+0.0057}$ & \textemdash & \textemdash & \textemdash & \textemdash & \textemdash & \textemdash & $-0.840_{-0.053}^{+0.054}$ & $-0.582_{-0.198}^{+0.207}$ \\
				CMB+DESI+Union3 & $2803.235$ & $66.00_{-0.79}^{+0.80}$ & $0.3271_{-0.0080}^{+0.0083}$ & \textemdash & \textemdash & \textemdash & \textemdash & \textemdash & \textemdash & $-0.672_{-0.084}^{+0.083}$ & $-1.048_{-0.274}^{+0.275}$ \\
				CMB+DESI+DES-SN & $4412.020$ & $67.50_{-0.54}^{+0.55}$ & $0.3121_{-0.0055}^{+0.0053}$ & \textemdash & \textemdash & \textemdash & \textemdash & \textemdash & \textemdash & $-0.811_{-0.054}^{+0.055}$ & $-0.700_{-0.224}^{+0.220}$ \\
				\midrule
				\multicolumn{12}{l}{\textbf{$\Lambda$CDM cosmology}} \\
				CMB+DESI+PP & $4193.871$ & $68.33_{-0.29}^{+0.28}$ & $0.3020_{-0.0036}^{+0.0037}$ & \textemdash & \textemdash & \textemdash & \textemdash & \textemdash & \textemdash & \textemdash & \textemdash \\
				CMB+DESI+Union3 & $2816.720$ & $68.30_{-0.29}^{+0.29}$ & $0.3024_{-0.0038}^{+0.0036}$ & \textemdash & \textemdash & \textemdash & \textemdash & \textemdash & \textemdash & \textemdash & \textemdash \\
				CMB+DESI+DES-SN & $4424.091$ & $68.29_{-0.27}^{+0.27}$ & $0.3025_{-0.0035}^{+0.0034}$ & \textemdash & \textemdash & \textemdash & \textemdash & \textemdash & \textemdash & \textemdash & \textemdash \\
				\bottomrule
				\bottomrule
			\end{tabular}%
		}
		\vspace{0.2cm}
		\resizebox{\textwidth}{!}{%
			\setlength{\tabcolsep}{3.0pt}
			\begin{tabular}{lccccccccc}
				\multicolumn{10}{c}{{\centering \bf Primordial and clustering parameters}}\\
				\toprule
				\toprule
				Dataset & $100\theta_s$ & $\Omega_{\rm b}h^2$ & $\Omega_{\rm c}h^2$ & $r_d$ & $\ln(10^{10}A_s)$ & $n_s$ & $\tau_{\rm reio}$ & $\sigma_8$ & $S_8$ \\
				\midrule
				\multicolumn{10}{l}{\textbf{BellDE cosmology}} \\
				CMB+DESI+PP & $1.04192_{-0.00028}^{+0.00028}$ & $0.02241_{-0.00014}^{+0.00014}$ & $0.11927_{-0.00095}^{+0.00095}$ & $147.25_{-0.23}^{+0.23}$ & $3.043_{-0.015}^{+0.015}$ & $0.9665_{-0.0036}^{+0.0038}$ & $0.054_{-0.007}^{+0.008}$ & $0.8123_{-0.0095}^{+0.0096}$ & $0.8262_{-0.0101}^{+0.0102}$ \\
				CMB+DESI+Union3 & $1.04191_{-0.00028}^{+0.00029}$ & $0.02240_{-0.00013}^{+0.00013}$ & $0.11940_{-0.00090}^{+0.00090}$ & $147.22_{-0.22}^{+0.22}$ & $3.042_{-0.014}^{+0.014}$ & $0.9662_{-0.0038}^{+0.0038}$ & $0.054_{-0.007}^{+0.007}$ & $0.8046_{-0.0097}^{+0.0098}$ & $0.8310_{-0.0099}^{+0.0096}$ \\
				CMB+DESI+DES-SN & $1.04192_{-0.00028}^{+0.00027}$ & $0.02241_{-0.00014}^{+0.00014}$ & $0.11926_{-0.00094}^{+0.00093}$ & $147.25_{-0.23}^{+0.23}$ & $3.043_{-0.014}^{+0.015}$ & $0.9662_{-0.0037}^{+0.0037}$ & $0.055_{-0.007}^{+0.007}$ & $0.8118_{-0.0090}^{+0.0092}$ & $0.8267_{-0.0098}^{+0.0099}$ \\
				\midrule
				\multicolumn{10}{l}{\textbf{SO cosmology}} \\
				CMB+DESI+PP & $1.04206_{-0.00027}^{+0.00028}$ & $0.02251_{-0.00012}^{+0.00012}$ & $0.11796_{-0.00066}^{+0.00066}$ & $147.49_{-0.19}^{+0.19}$ & $3.051_{-0.015}^{+0.015}$ & $0.9698_{-0.0034}^{+0.0034}$ & $0.059_{-0.007}^{+0.007}$ & $0.8045_{-0.0075}^{+0.0075}$ & $0.8137_{-0.0089}^{+0.0090}$ \\
				CMB+DESI+Union3 & $1.04201_{-0.00028}^{+0.00027}$ & $0.02247_{-0.00013}^{+0.00013}$ & $0.11860_{-0.00072}^{+0.00072}$ & $147.37_{-0.20}^{+0.20}$ & $3.047_{-0.014}^{+0.014}$ & $0.9683_{-0.0034}^{+0.0034}$ & $0.057_{-0.007}^{+0.007}$ & $0.7945_{-0.0078}^{+0.0078}$ & $0.8257_{-0.0097}^{+0.0095}$ \\
				CMB+DESI+DES-SN & $1.04207_{-0.00027}^{+0.00027}$ & $0.02251_{-0.00012}^{+0.00013}$ & $0.11796_{-0.00067}^{+0.00068}$ & $147.49_{-0.20}^{+0.20}$ & $3.051_{-0.015}^{+0.014}$ & $0.9697_{-0.0034}^{+0.0034}$ & $0.060_{-0.007}^{+0.007}$ & $0.8048_{-0.0073}^{+0.0073}$ & $0.8137_{-0.0091}^{+0.0090}$ \\
				\midrule
				\multicolumn{10}{l}{\textbf{CPL cosmology}} \\
				CMB+DESI+PP & $1.04194_{-0.00026}^{+0.00027}$ & $0.02244_{-0.00013}^{+0.00013}$ & $0.11893_{-0.00087}^{+0.00091}$ & $147.31_{-0.22}^{+0.22}$ & $3.044_{-0.014}^{+0.014}$ & $0.9669_{-0.0037}^{+0.0036}$ & $0.055_{-0.007}^{+0.007}$ & $0.8089_{-0.0089}^{+0.0092}$ & $0.8233_{-0.0097}^{+0.0096}$ \\
				CMB+DESI+Union3 & $1.04192_{-0.00029}^{+0.00029}$ & $0.02240_{-0.00013}^{+0.00013}$ & $0.11938_{-0.00090}^{+0.00085}$ & $147.23_{-0.20}^{+0.21}$ & $3.041_{-0.014}^{+0.015}$ & $0.9661_{-0.0038}^{+0.0038}$ & $0.054_{-0.008}^{+0.008}$ & $0.7979_{-0.0102}^{+0.0104}$ & $0.8331_{-0.0096}^{+0.0101}$ \\
				CMB+DESI+DES-SN & $1.04192_{-0.00029}^{+0.00028}$ & $0.02241_{-0.00014}^{+0.00014}$ & $0.11915_{-0.00086}^{+0.00090}$ & $147.28_{-0.22}^{+0.22}$ & $3.043_{-0.015}^{+0.015}$ & $0.9667_{-0.0036}^{+0.0035}$ & $0.054_{-0.007}^{+0.008}$ & $0.8096_{-0.0086}^{+0.0090}$ & $0.8257_{-0.0095}^{+0.0101}$ \\
				\midrule
				\multicolumn{10}{l}{\textbf{\boldmath$\Lambda$CDM cosmology}} \\
				CMB+DESI+PP & $1.04207_{-0.00027}^{+0.00026}$ & $0.02251_{-0.00012}^{+0.00012}$ & $0.11784_{-0.00065}^{+0.00063}$ & $147.52_{-0.19}^{+0.19}$ & $3.052_{-0.014}^{+0.014}$ & $0.9698_{-0.0034}^{+0.0035}$ & $0.060_{-0.007}^{+0.007}$ & $0.8079_{-0.0060}^{+0.0060}$ & $0.8106_{-0.0082}^{+0.0084}$ \\
				CMB+DESI+Union3 & $1.04207_{-0.00026}^{+0.00028}$ & $0.02251_{-0.00012}^{+0.00012}$ & $0.11791_{-0.00068}^{+0.00066}$ & $147.50_{-0.19}^{+0.20}$ & $3.051_{-0.014}^{+0.015}$ & $0.9695_{-0.0033}^{+0.0032}$ & $0.060_{-0.007}^{+0.007}$ & $0.8076_{-0.0057}^{+0.0061}$ & $0.8108_{-0.0081}^{+0.0082}$ \\
				CMB+DESI+DES-SN & $1.04208_{-0.00028}^{+0.00027}$ & $0.02251_{-0.00012}^{+0.00012}$ & $0.11794_{-0.00062}^{+0.00060}$ & $147.49_{-0.19}^{+0.19}$ & $3.051_{-0.015}^{+0.015}$ & $0.9696_{-0.0032}^{+0.0031}$ & $0.060_{-0.007}^{+0.007}$ & $0.8078_{-0.0062}^{+0.0062}$ & $0.8112_{-0.0085}^{+0.0082}$ \\
				\bottomrule
				\bottomrule
			\end{tabular}%
		}
		\caption{
			Cosmological constraints for the BellDE, SO, CPL and flat $\Lambda$CDM models. All parameter constraints are posterior means with central $68\%$ credible intervals.
		}
		\label{tab:combined_cosmo_constraints_68_fourmodel}
	\end{table*}
	
	\subsubsection{BellDE}
	\label{subsubsec:Bell_results}
	
\begin{figure*}[t]
		\begin{tabular}{cc}
			{\bf BellDE} & {\bf SO}\\
			\includegraphics[scale=0.45]{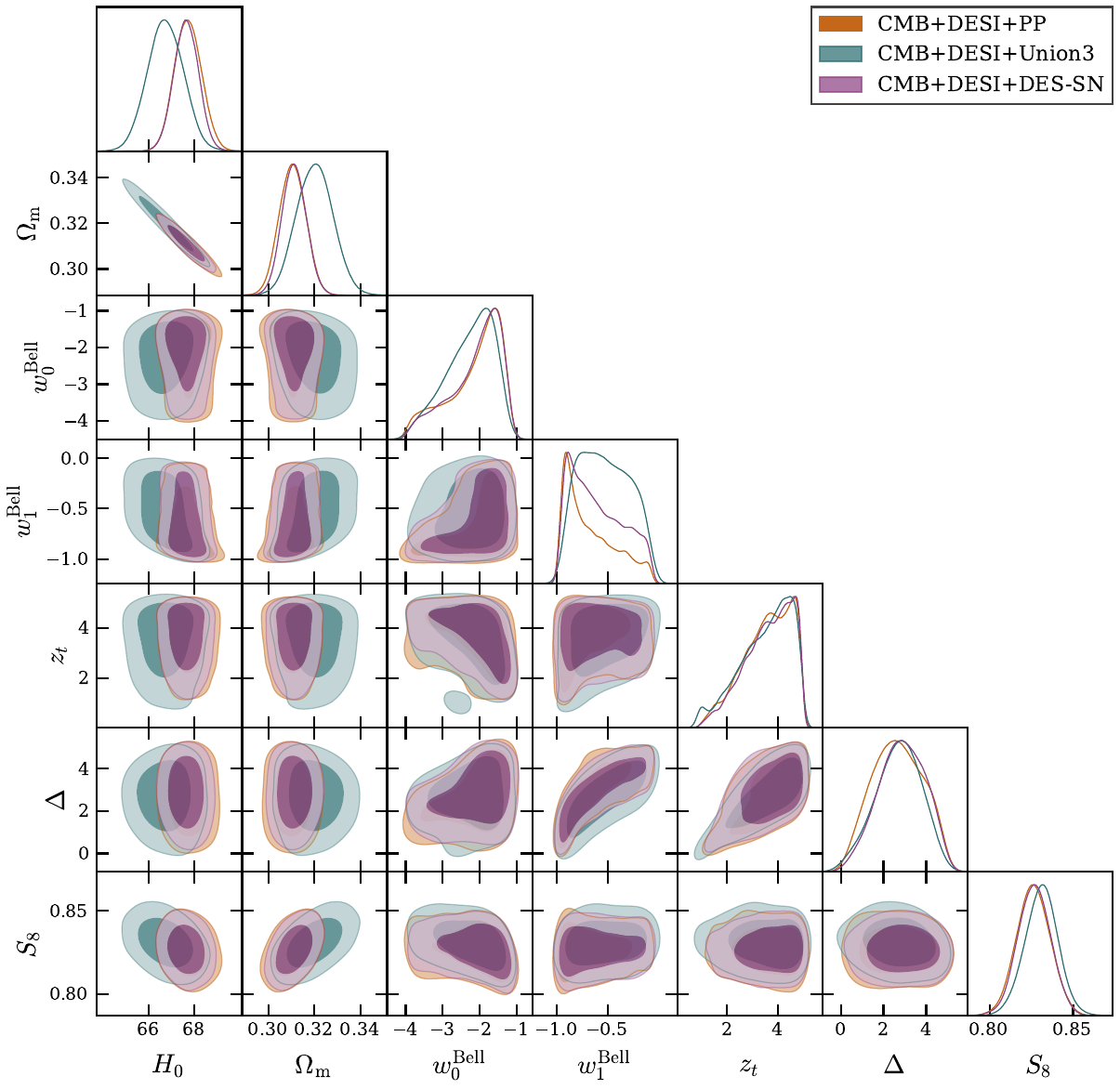} &
			\includegraphics[scale=0.45]{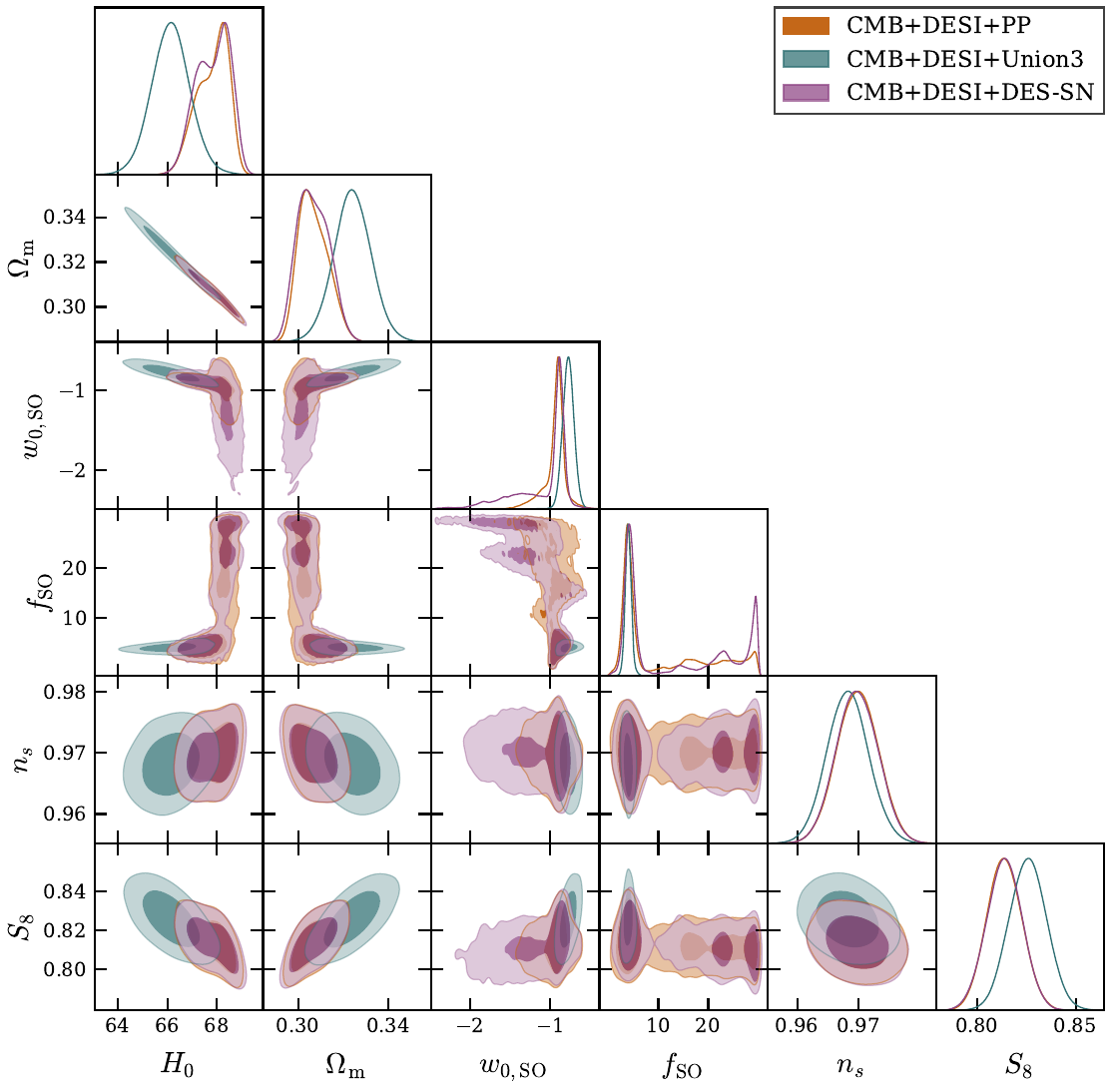}
		\end{tabular}
		\caption{Marginalized posterior distributions of BellDE and SO from the PP, Union3 and DES-SN combinations. Inner and outer contours denote the $68\%$ and $95\%$ credible regions.}
		\label{fig:Bell_posterior}
		\label{fig:SO_posterior}
	\end{figure*}

\begin{figure*}[t]	\includegraphics[scale=0.5]{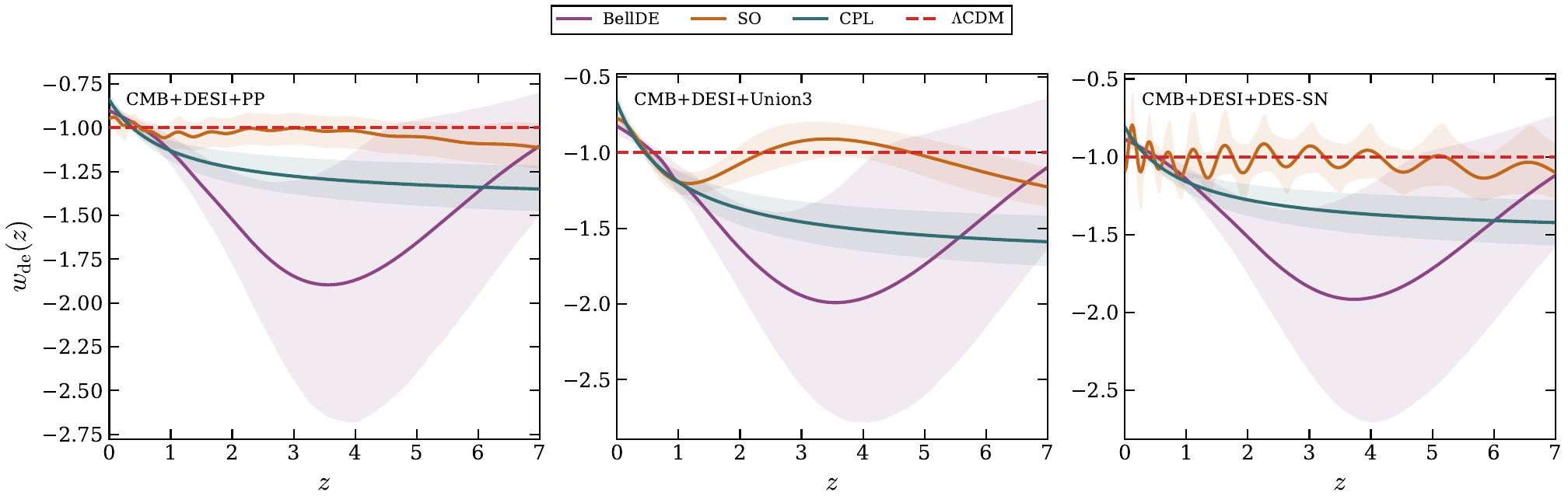}
		\caption{Reconstructed $w_{\rm de}(z)$ for BellDE, SO, CPL and $\Lambda$CDM for the PP, Union3 and DES-SN combinations. Curves denote posterior means and bands the central $68\%$ credible regions, the horizontal dashed line marks $w_{\rm de}=-1$.}
		\label{fig:eos_reconstruction}
	\end{figure*}

	The marginalized posterior distributions of the parameters are shown in Fig.~\ref{fig:Bell_posterior}. For the BASE+PP compilation, the model yields $H_0=67.72^{+0.60}_{-0.59}\,{\rm km\,s^{-1}\,Mpc^{-1}}$ and $\Omega_{\rm m}=0.3104^{+0.0057}_{-0.0058}$. The EOS parameters take the values $w_0^{\rm Bell}=-2.142^{+0.712}_{-0.874}$ and $w_1^{\rm Bell}=-0.671^{+0.307}_{-0.249}$, with the transition redshift $z_t=3.584^{+1.024}_{-1.029}$ and $\Delta=2.674^{+1.358}_{-1.282}$. Due to the large negative value of $w_0^{\rm Bell}$, the model exhibits a stronger phantom behavior than the CPL parametrization, while asymptotically approaching the quintessence regime at high redshifts and in the future, as shown in Fig.~\ref{fig:eos_reconstruction}. The model yields a present-day EOS value very close to that of CPL, $\sim-0.840$. The posterior distributions show that $z_t$ and $\Delta$ are positively correlated, indicating a degeneracy between these parameters. Owing to this degeneracy, the transition from the phantom to the quintessence regime occurs at a very high redshift. Consequently, the CPL evolution of the equation of state becomes a subset of the BellDE evolution at intermediate redshifts. Thus, despite the additional degrees of freedom in the model, it does not exhibit a transition to the quintessence regime within the interval $0.5<z<3.0$.
	
	For the Union3 compilation, the model yields a comparatively lower value of $H_0\simeq66.7\,{\rm km\,s^{-1}\,Mpc^{-1}}$ and a larger value of $\Omega_{\rm m}\simeq0.320$, with $w_1^{\rm Bell}\sim-0.527$, while the remaining parameters are very close to those obtained for the PP compilation. However, for the DES-SN compilation, the model exhibits a trend similar to that of the Pantheon+ compilation. Thus, for both of these samples, the EOS evolution follows a similar trend, as shown in Fig.~\ref{fig:eos_reconstruction}.
	
	\subsubsection{Sourced Oscillation model}
	\label{subsubsec:SO_results}
	
\begin{figure*}[t]
		\begin{tabular}{ccc}
			{\bf PP} & {\bf Union3} & {\bf DES-SN}\\
			\includegraphics[scale=0.35]{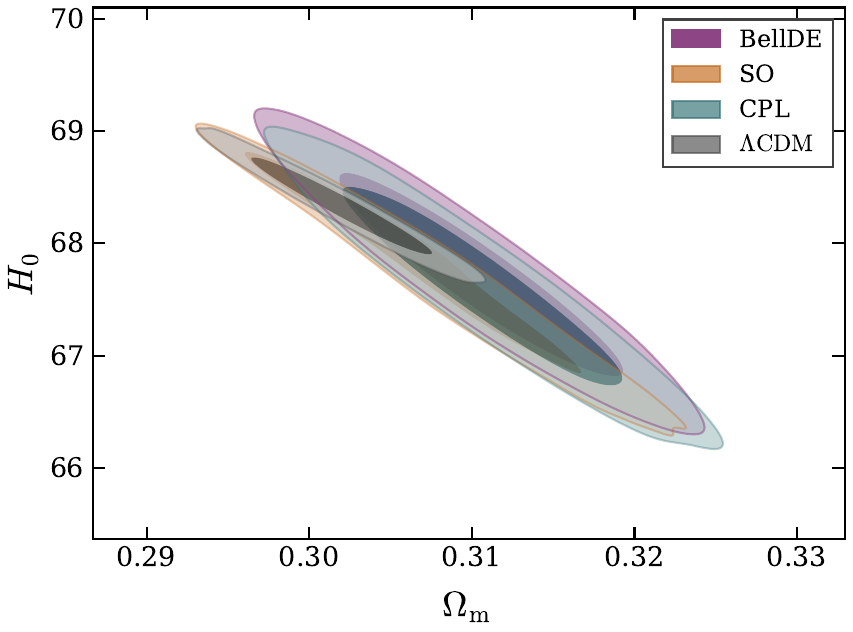}&
			\includegraphics[scale=0.35]{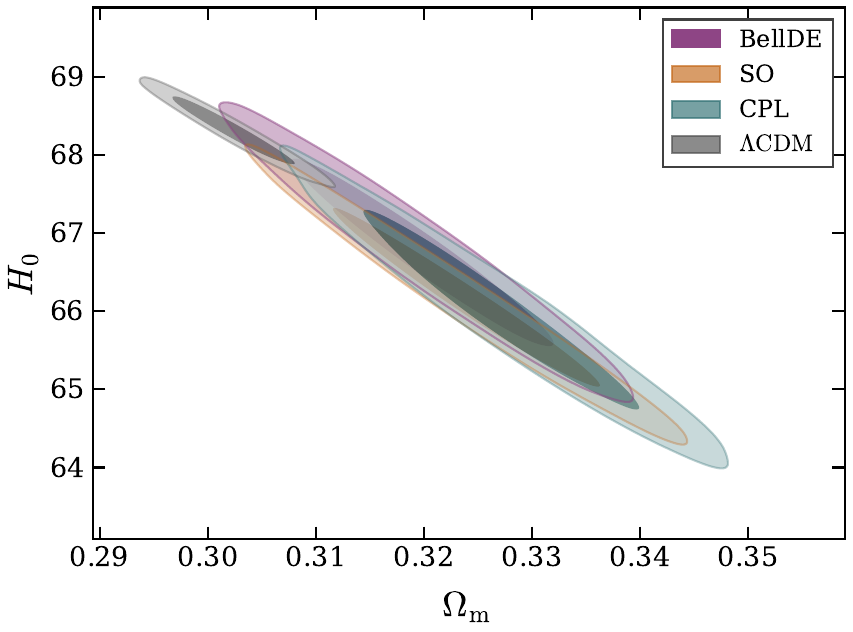}&
			\includegraphics[scale=0.35]{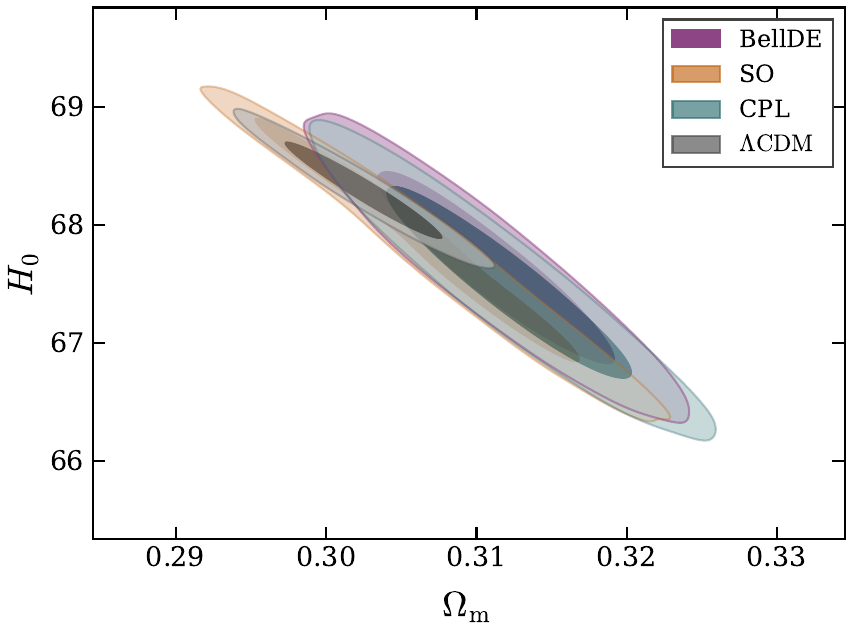}
		\end{tabular}
		\caption{Marginalized 2D posterior distributions of $H_0 -\Omega_{m0}$ for BellDE, SO, CPL and $\Lambda$CDM.}
		\label{fig:omho_2d}
	\end{figure*}

	The marginalized posterior distributions of the parameters are shown in Fig.~\ref{fig:SO_posterior}, where the EOS parameter $w_{0,\rm SO}$ remains close to $-1$ for all the supernova compilations. The parameter constraints are tabulated in Tab.~\ref{tab:combined_cosmo_constraints_68_fourmodel}. The central value of the EOS parameter, $w_{0,\rm SO}=-1.089_{-0.376}^{+0.242}$, lies in the phantom regime for the DES-SN dataset, whereas for the other samples, such as Pantheon+, the central value $w_{0,\rm SO}=-0.945^{+0.096}_{-0.111}$ lies in the quintessence regime. This demonstrates the differences in the geometric constraints arising from the distinct supernova compilations. For both PP and DES-SN, the frequency is $f_{\rm SO}\sim12$, whereas for Union3, $f_{\rm SO}\sim4$. For all compilations, the dominant peak in the frequency distribution occurs at $f_{\rm SO}<10$. However, only for DES-SN and PP, we obtain a secondary peak at $f_{\rm SO}>25$, resulting in higher preferred frequencies for both samples. We stress, however, that the frequency posteriors are broad and bimodal: the present data constrain primarily the oscillation amplitude through $w_{0,\rm SO}$, rather than the oscillation period, so the preferred values of $f_{\rm SO}$ should be interpreted with caution. Definite constraints on $f_{\rm SO}$ will require tighter geometric measurements over $0.5\lesssim z\lesssim2$, as expected from the upcoming Euclid, and LSST. Owing to this spread, the equation of state shifts toward more negative values, whereas for Union3, $w_{0,\rm SO}=-0.778\pm0.069$ remains in the quintessence regime. The model yields values of $H_0$ and $\Omega_{\rm m0}$ very similar to those of BellDE and CPL for Union3, whereas for PP and DES-SN, the posterior distributions closely follow those of $\Lambda$CDM, as shown in Fig.~\ref{fig:omho_2d}.
	
	The reconstructed EOS for this model is shown in Fig.~\ref{fig:eos_reconstruction}. For the PP sample, the mean equation of state crosses above $-1$ around $z=0.5$, yielding a value very close to those of CPL and BellDE. The oscillation amplitude is very small, and the equation of state falls below $-1$ over $0.5<z<2$. It then approaches $-1$ over $2<z<4$ before moving towards more negative values at higher redshifts ($z>4$). The 68\% credible band remains very narrow and close to $-1$, indicating tighter constraints than those obtained for BellDE and CPL. For the Union3 compilation, the lower frequency results in oscillatory peaks near $z\sim1$ and $z\sim3$, with crossings of the $w_{\rm de}=-1$ line around $z\simeq0.5$, $2.5$, and $4.5$. Consequently, the model remains in the quintessence regime for $z\in[0,0.5]$, transitions to the phantom regime for $z\in[0.5,2.5]$, returns to the quintessence regime for $z\in[2.5,4.5]$, and moves towards more negative phantom values beyond $z>6.0$. Thus, for this dataset, the reconstruction clearly reveals a tension between CPL and SO in the range $2.5<z<6.0$. Finally, for the DES-SN compilation, the model exhibits higher-frequency oscillations with larger amplitudes around $-1$. These frequent oscillations arise from the model's sensitivity to the distinct geometric constraints associated with the different supernova compilations.
	
\begin{figure*}
		\includegraphics[scale=0.5]{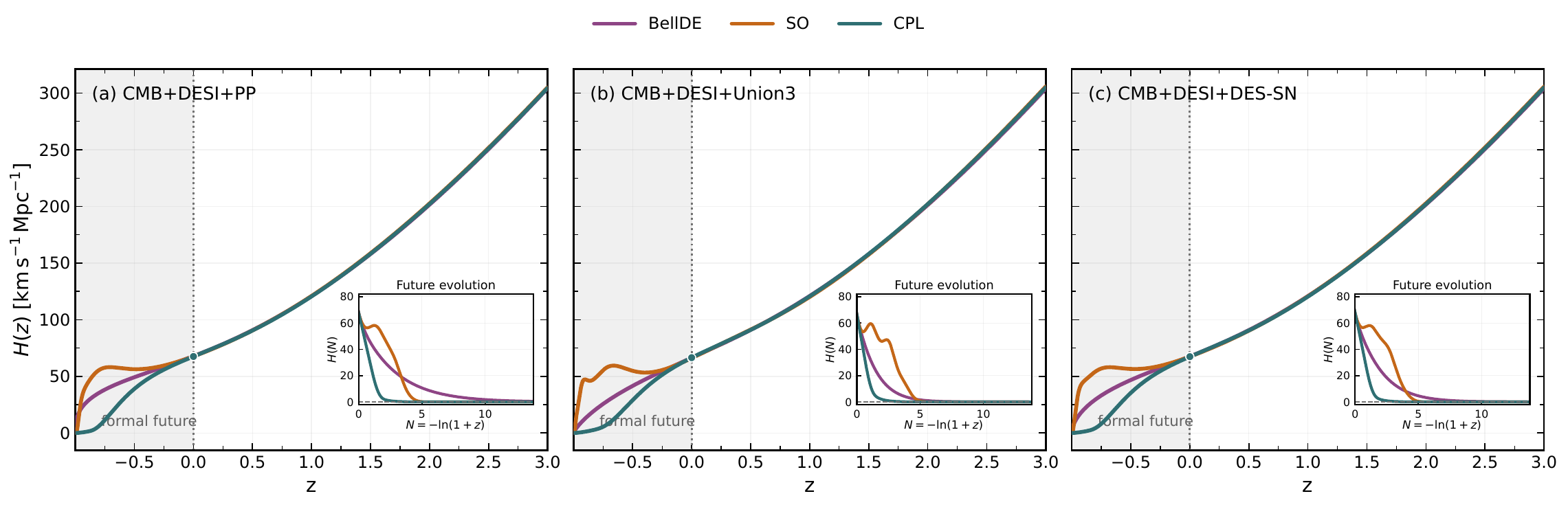}
		\caption{The Hubble evolution plot corresponding to the dynamical models of the dark energy.}
		\label{fig:hubble_evo}
	\end{figure*}

	We also show the marginalized posterior distributions for CPL and $\Lambda$CDM in Fig.~\ref{fig:lcdm_cpl_posterior}. The Hubble evolution for each model, corresponding to the parameter constraints, is plotted in Fig.~\ref{fig:hubble_evo}. At higher redshifts, all models trace similar evolution. However, when extrapolated into the future, the equation of state settles into the $w_{\rm de}>-1$ regime, the dark energy density dilutes, and the Hubble parameter decreases asymptotically towards zero, rendering decelerated expansion approaching a static state. For CPL, the energy density dilutes rapidly, whereas for the SO model, the Universe remains in the accelerating phase for a brief interval of e-folds, $0<N<5$. Once the oscillation mean shifts above $w_{\rm de}=-1$, the dark energy density is no longer nearly constant and the expansion decelerates permanently. Thus, all the dynamical models predict a transient epoch of cosmic acceleration followed by an asymptotically static Universe, in contrast to the eternal de Sitter expansion of the standard model.
	
\begin{figure}[t]
		\includegraphics[scale=0.25]{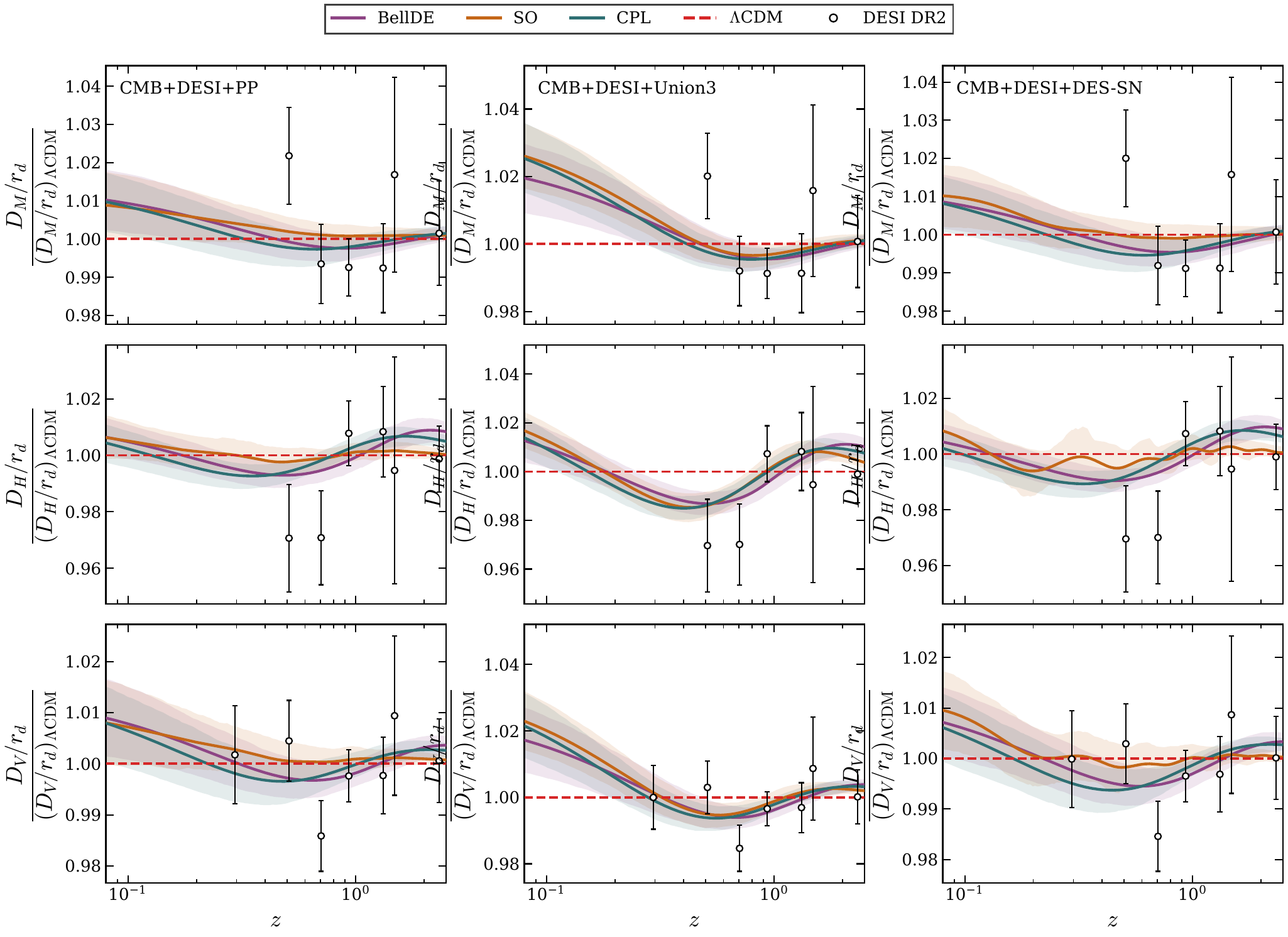}
		\caption{DESI DR2 distance reconstructions for BellDE, SO and CPL relative to the corresponding $\Lambda$CDM prediction. Rows show $D_M/r_d$, $D_H/r_d$ and $D_V/r_d$, while columns correspond to PP, Union3 and DES-SN. Curves and bands denote posterior means and central 68\% credible regions, black points show the DESI DR2 BAO measurements.}
		\label{fig:bao_relative}
	\end{figure}

	Additionally, we illustrate the ratios of the geometric distances, including $D_{M}/r_d$, $D_{H}/r_d$, and $D_{V}/r_d$, relative to the fiducial model in Fig.~\ref{fig:bao_relative}. The values of $r_d$ for all models are taken from Tab.~\ref{tab:combined_cosmo_constraints_68_fourmodel}. All dynamical models behave very similarly across the redshift range for each dataset, although slight deviations can be observed for the SO model.
	
\begin{figure*}[!t]
		\centering
		
		\resizebox{0.9\textwidth}{!}{%
			\includegraphics{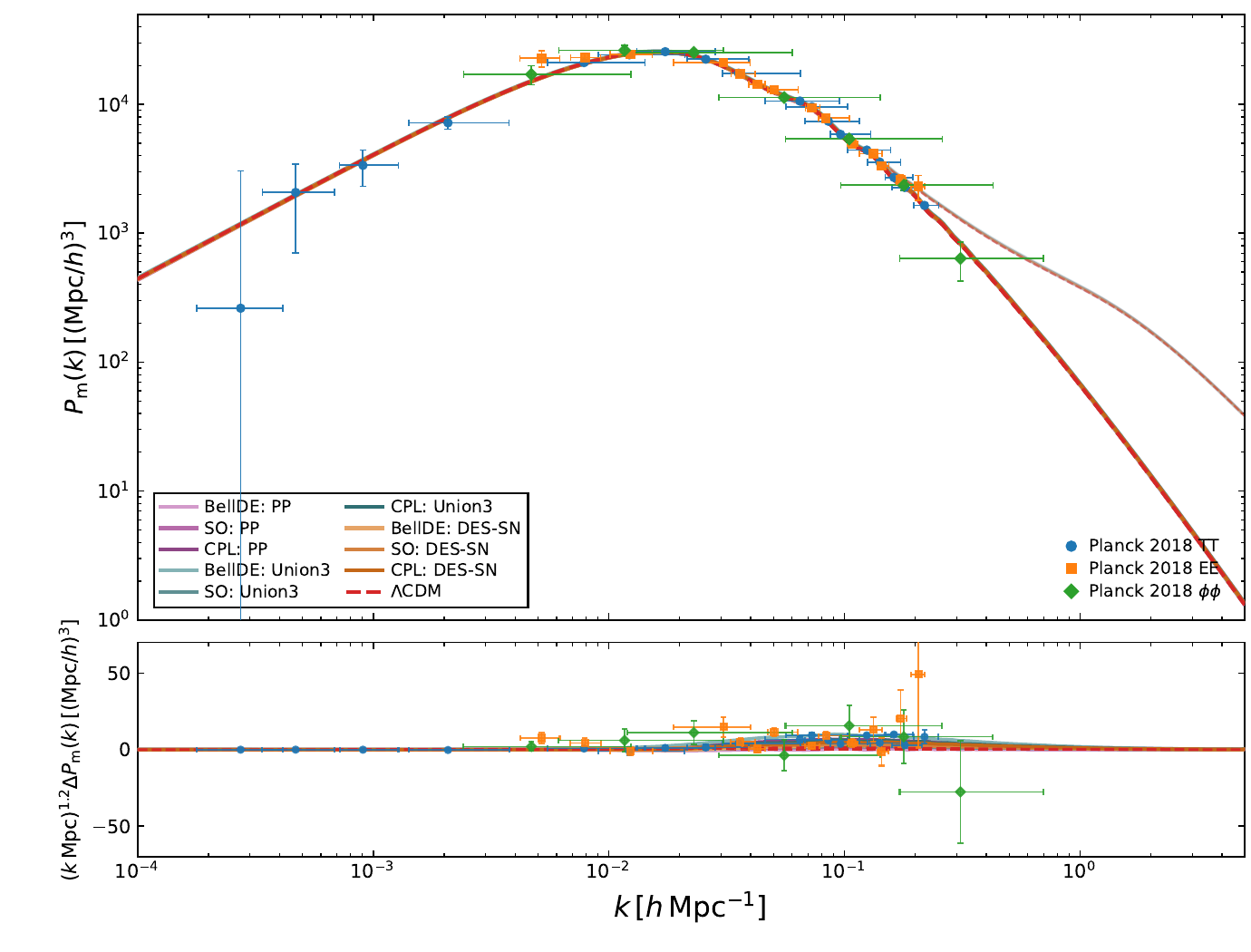}
			\includegraphics{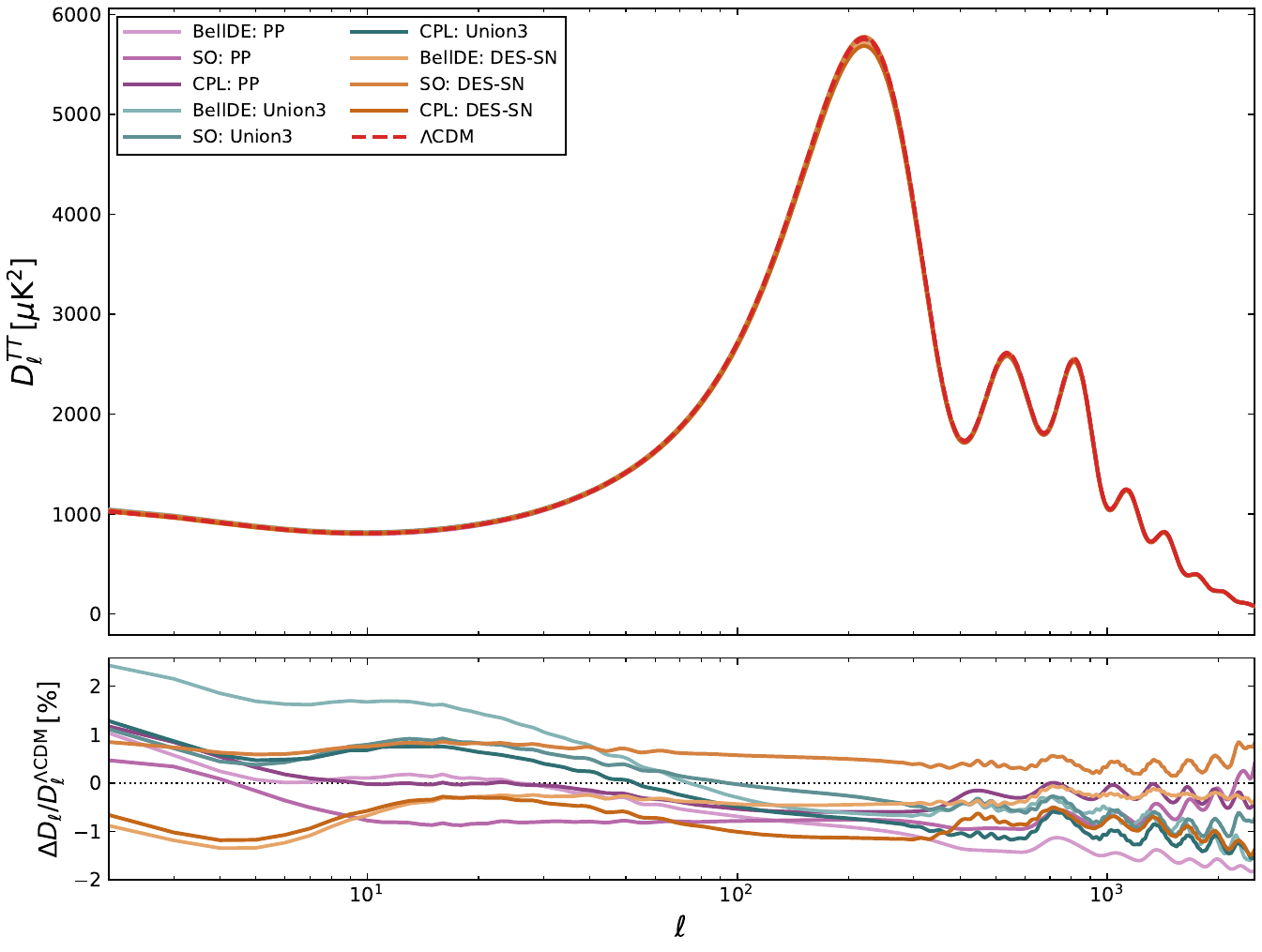}
		}
		
		\caption{Left: joint matter power spectra and scaled residuals for the three data combinations. Points show the common Planck TT, EE and lensing sensitivities. Right: best fitting lensed CMB temperature spectra and fractional residuals relative to the corresponding $\Lambda$CDM fits.
		}
		\label{fig:cmb_matter_power}
	\end{figure*}

	We also plot the matter power spectra at $z=0$ for different wavenumbers $k$ in Fig.~\ref{fig:cmb_matter_power}. The difference in the matter power spectrum, $\Delta P_{\rm m}(k)$, reveals only a minuscule difference between $\Lambda$CDM and the other dynamical models. Thus, the models exhibit negligible differences on CMB scales, which is also evident in the CMB angular power spectrum $D_{\ell}^{\rm TT}$. The differences are approximately $\sim1\%$ for each dynamical model.
	
	\subsection{Statistical Metric}
	
	To evaluate the viability of each model, we compute its log-evidence relative to the baseline model as
	\begin{equation}
		\Delta\log Z_i = \log Z_i - \log Z_{\Lambda{\rm CDM}} \ .
		\label{eq:model_evidence}
	\end{equation}
	We adopt the model-selection criteria given in Refs.~\cite{Speagle:2019ivv,Kass:1995loi}, where $0 < |\Delta \log Z| < 1$ indicates inconclusive evidence, $1 < |\Delta \log Z| < 2.5$ corresponds to weak evidence, $2.5 < |\Delta \log Z| < 5$ signifies moderate evidence, and $|\Delta \log Z| > 5$ represents strong evidence. The evidence for each model is tabulated in Tab.~\ref{tab:model_evidence}. According to this scale, BellDE is weakly preferred over $\Lambda$CDM for the PP combination ($\Delta\log Z=+2.14$) and moderately preferred for both Union3 ($+3.65$) and DES-SN ($+3.38$). The SO model is inconclusive for PP ($-0.05$) and Union3 ($+0.29$) and weakly preferred for DES-SN ($+1.59$), while CPL is weakly preferred for Union3 ($+2.16$), weakly disfavored for PP ($-1.85$), and inconclusive for DES-SN ($+0.02$). Nevertheless, we note that BellDE has a higher number of degrees of freedom, leading to a broader spread around the mean equation of state at higher redshifts, as shown in Fig.~\ref{fig:eos_reconstruction}. In contrast, SO exhibits tighter constraints, with the equation of state remaining very close to $-1$ around low redshift. Thus, SO serves as a potentially viable alternative to CPL.

	\begin{table}[t]
		\centering
		\footnotesize
		\renewcommand{\arraystretch}{1.2}
		\setlength{\tabcolsep}{8pt}
		\begin{tabular}{@{}llcc@{}}
			\toprule
			\toprule
			\textbf{Dataset} & \textbf{Model} & \(\boldsymbol{\log Z}\) & \(\boldsymbol{\Delta \log Z}\) \\
			\midrule
			
			\multirow{4}{*}{CMB+DESI+PP} 
			& \textbf{BellDE}     & \(-2148.6785\) & \(+2.1370\) \\
			& \textbf{SO}         & \(-2150.8675\) & \(-0.0520\) \\
			& \textbf{CPL}        & \(-2152.6628\) & \(-1.8473\) \\
			& \textbf{$\Lambda$CDM} & \(-2150.8155\) & \(0\) \\
			\midrule
			
			\multirow{4}{*}{CMB+DESI+Union3} 
			& \textbf{BellDE}     & \(-1458.3268\) & \(+3.6490\) \\
			& \textbf{SO}         & \(-1461.6886\) & \(+0.2872\) \\
			& \textbf{CPL}        & \(-1459.8134\) & \(+2.1624\) \\
			& \textbf{$\Lambda$CDM} & \(-1461.9758\) & \(0\) \\
			\midrule
			
			\multirow{4}{*}{CMB+DESI+DES-SN} 
			& \textbf{BellDE}     & \(-2262.2301\) & \(+3.3784\) \\
			& \textbf{SO}         & \(-2264.0199\) & \(+1.5886\) \\
			& \textbf{CPL}        & \(-2265.5891\) & \(+0.0194\) \\
			& \textbf{\boldmath$\Lambda$CDM} & \(-2265.6085\) & \(0\) \\
			
			\bottomrule
			\bottomrule
		\end{tabular}
		\caption{
			Bayesian evidence for the BellDE, SO, CPL, and $\Lambda$CDM models across the three data combinations. We report the natural logarithm of the evidence (\(\log Z\)) and the difference relative to $\Lambda$CDM, defined as \(\Delta \log Z \equiv \log Z_{\rm model} - \log Z_{\Lambda{\rm CDM}}\). A negative \(\Delta \log Z\) indicates a preference for $\Lambda$CDM.
		}
		\label{tab:model_evidence}
	\end{table}

	\section{Conclusions}
	\label{sec:conclusions}
	
	In this work, we investigated late time dark energy models by parameterizing their equation of state $w_{\rm de}$. We constructed a novel equation of state that satisfies a simple harmonic oscillator differential equation with a source term proportional to the cube of the number of e-folds, $N$. We set the proportionality constant equal to the present day matter energy density parameter, $\Omega_{\rm m0}$. Owing to its specific functional form, the model has only two free parameters: $w_{0}$ and the frequency $f$. We then constrained the model parameters by comparing its predictions with various combinations of observational data, including Planck 2018, DESI DR2, and three distinct supernova compilations: Pantheon+, Union3 and DES-SN. We adopted Planck and DESI as the baseline datasets and incorporated each supernova compilation separately. We compared the sourced oscillatory (SO) model with BellDE, CPL and flat $\Lambda$CDM. Unlike the CPL model, which is not generally well suited to extrapolation into the far future, both SO and BellDE can be extrapolated beyond the present epoch without this particular limitation. Furthermore, the SO model has only two free parameters and is dynamically driven. It also yields tighter parameter constraints than both CPL and BellDE, making it a competitive alternative with richer dynamical behavior. The model exhibits dynamics very close to those of $w_{\rm de}=-1$ over the redshift range $0.6<z<3.0$ for Pantheon+ and DES-SN, with deviations emerging at $z<0.6$ and at higher redshifts. The log-evidence for the SO model ranges from inconclusive (Pantheon+, Union3) to weak (DES-SN) relative to $\Lambda$CDM, and is generally weaker than that of CPL, while BellDE is moderately preferred for Union3 and DES-SN and weakly preferred for Pantheon+. We also note that none of the dynamical models alleviates the Hubble tension, as all of them yield lower $H_0$ values than $\Lambda$CDM. When extrapolated into the future, all the dynamical models predict that the present accelerated expansion is transient: the dark energy density dilutes, the expansion decelerates permanently, and the Hubble parameter approaches zero as scale factor approaches $(a \sim e^{5})$, leaving an asymptotically static Universe. The SO model extends the accelerated epoch by a few e-folds relative to CPL before reaching the same asymptotic regime. Several limitations of the present analysis should be kept in mind: the normalization of the source term is fixed by convention, the oscillation frequency remains weakly constrained and exhibits a bimodal posterior for two of the three supernova compilations, and a microscopic realization of the multiple crossing dynamics would require either non-minimal coupling of the field to curvature or interactions within the dark sector, both of which lie beyond the background level description adopted here. Whether forthcoming data from the full DESI survey, Euclid and LSST can constrain a definite oscillation frequency remains an open question. Answering it would turn the transient acceleration scenario into a falsifiable prediction.

	\begin{acknowledgments}
		This work is supported by the National Key R\&D Program of China (Grant Nos. 2024YFA1611503 and 2020YFC2201503). S.H. acknowledges the support of the National Natural Science Foundation of China under Grants No.~12275238, No.~12542053, and No.~W2433018. This work is also partly supported by the Urumqi Nanshan Astronomy and Deep Space Exploration Observation and Research Station of Xinjiang (XJYWZ2303), the Central Guidance for Local Science and Technology Development Fund (Grant No. ZYYD2026JD01), and the Astrometric Reference Frame project (Grant No. JZZX-0102). 
        
	\end{acknowledgments}
	
	\bibliographystyle{apsrev4-2}
	\bibliography{references}
	
\end{document}